\documentclass[12pt]{iopart}
\usepackage{graphicx}
\renewcommand{\br}{{\bf r}}

\newcommand{\ux}{\hat{\bf x}}
\newcommand{\uy}{\hat{\bf y}}

\begin{document}
\paper[Transport in the Hamiltonian XY-model]
{A simulation study of energy transport in the Hamiltonian XY-model}

\author{Luca Delfini \dag, Stefano Lepri \dag  
\footnote[4]{To
whom correspondence should be addressed (lepri@inoa.it)}
and Roberto Livi \ddag
}
\address{\dag\ Istituto Nazionale di Ottica Applicata and\\
Istituto dei Sistemi Complessi, Consiglio Nazionale delle Ricerche, 
Sezione Territoriale di Firenze, largo E. Fermi 6
I-50125 Firenze, Italy}
\address{\ddag Dipartimento di Fisica, Universit\`a di Firenze, 
Sezione INFN, Unit\'a INFM and CSDC Firenze,
via G. Sansone 1 I-50019, Sesto Fiorentino, Italy }. 

\begin{abstract}
The transport properties of the planar rotator model on a square lattice are
analyzed by means of microcanonical and non--equilibrium  simulations. Well 
below the Kosterlitz--Thouless--Berezinskii transition temperature, both
approaches  consistently indicate that the  energy current autocorrelation
displays a long--time tail decaying as $t^{-1}$. This yields a thermal
conductivity coefficient which diverges logarithmically with the lattice size.
Conversely, conductivity is found to be finite in the high--temperature
disordered phase. Simulations close to the transition temperature are 
insted limited by slow convergence that is presumably due to the 
slow kinetics of vortex pairs.   
\end{abstract}

\submitto{Journal of Statistical Mechanics: theory and experiment}

\maketitle

\section{Introduction}

Physical phenomena in reduced spatial dimension ($d=1,2$) are often
qualitatively different from their three-dimensional counterparts.  The
overwhelming role of statistical fluctuations and the presence of constraints
in the motion of excitations can lead to peculiar  effects like the
impossibility  of long--range order. In the context of non--equilibrium
statistical mechanics, the existence of long--time tails in fluids \cite{PR75}
leads to ill--defined transport coefficients, thus implying a breakdown of the
phenomenological constitutive laws of hydrodynamics \cite{kirk}. 

A remarkable example is
the anomalous behavior of heat conductivity for $d\le 2$. This issue attracted
a renovated interest within the statistical--mechanics community  after the
discovery that the thermal conductivity of anharmonic chains diverge in the
thermodynamic limit~\cite{LLP97}. Since then, those anomalies have been 
detected in a series of different models. An exhaustive account is given in
Ref.~\cite{LLP03}, where the effects of lattice dimensionality, disorder and
external fields for the validity of Fourier's law are discussed in detail.  
The signature of anomalous behavior is a non--integrable algebraic decay 
of the correlator of the heat current ${\bf J}$ (the Green-Kubo 
integrand) at large times  
\begin{equation} 
\langle {\bf J}(t)\cdot {\bf J}(0)\rangle \; 
\propto \; t^{-(1-\alpha)}  \, ,\quad t\to +\infty  
\label{anomal}
\end{equation}
where $0\le \alpha < 1$ and  $\langle \,\, \rangle$ is the equilibrium average.
For a finite system of linear size $L$ this implies that the  finite-size
conductivity $\kappa(L)$ diverges in the $L\to \infty$ limit. In fact, in the
framework of linear-response theory, $\kappa$ can be estimated by
cutting--off the integral in the Green-Kubo formula at the transit time
$L/v$ ($v$ being some propagation velocity of energy carriers).
Taking into account Eq.~(\ref{anomal}) one straightforwardly obtains 
$\kappa \propto L^{\alpha}$. 

Simulation studies of specific models~\cite{LLP03} as well as analytic
arguments~\cite{NR02}, lead to the surmise that anomalous conductivity should
occur generically whenever momentum is conserved.  Moreover, the exponent
$\alpha$ should be largely independent on the  microscopic details as suggested
by a the renormalization--group  calculation of Ref.~\cite{NR02} that predicts
$\alpha = (2-d)/(2+d)$. For $d=1$ the resulting value $\alpha=1/3$ is roughly
close to the numerical estimates although some substantial deviations have been
observed in specific cases~\cite{D01,LLP03b,CP03}. The situation is even more
controversial  $d=2$ where the predicted $t^{-1}$ decay yields a
logarithmic singularity which is consistent with simulation data
\cite{LL00}, while other works report significative deviations and 
dimensional crossovers \cite{GY02}. 

When extending the analysis to the $2d$ case, one naturally wonders how the
possibility of observing critical phases may affect the anomalous energy
conduction. The first example one may think of is of course the Ising model. It
has however been shown that, at least for a specific choice of the  spin
dynamics, the latter displays a normal conductivity at all temperatures
\cite{saito}. In a more general perspective, this is consistent with the idea
the breakdown of momentum conservation removes  transport anomalies. Indeed,
the field--theoretic counterpart of the Ising model, the so called
$\phi^4$--theory, acquires an on--site non--linear interaction that breaks
translational invariance. Although we are not aware of any study of this model
in the ordered phase, it is known that the lattice $\phi^4$--model in the 
high--temperature phase displays a finite conductivity in the thermodynamic 
limit \cite{Hu}. 

In this present paper, we present a simulation study of
the transport properties of a model of rotators coupled on a square lattice,
akin to the celebrated XY--model (see e.g. \cite{gulacsi} and references
therein for a comprehensive review). As it is well known, the latter is
characterized by the presence of the so called Kosterlitz-Thouless-Berezinskii
(KTB) phase transition at finite temperature between a disordered
high--temperature phase and a low--temperature one, where vortexes condensate.

As recalled above, the fact the momentum (actually the angular momentum) is a
constant of motion, makes this model a candidate for observing anomalous
behavior. On the other hand, its $1d$ version is the only known exception to
this requirement and displays normal trasport, due to the presence  of
``dynamical defects" in the form of localized rotations that act  as scattering
centers for the heat carriers \cite{GLPV00}. On the basis of this observation
it is extremely interesting to investigate if and how the vortices play 
a similar role in the $2d$ case. 

Before entering the details of the present work, it is important to mention
that some evidence of the role of the vortex unbinding on the transport
properties of the (modified) XY--model have been reported in Ref.~\cite{DP97}.
Nonetheless, those results are mainly qualitative and we are thus motivated to
undertake a more detailed analysis.  

The paper is organized as follows. In section 2 we introduce the model and its microcanonical simulation. In Section 3 we recall the technique we used
to investigate the non-equilibrium stationary state. The outcomes of numerical
simulations for the disordered and critical phases are reported in Sections 4
and 5, respectively. Finally, we summarize and discuss our results in Section 6. 

\section{Hamiltonian dynamics of the XY model}

The XY or planar rotator model consists of a set of classical ``spins"   
${\bf S}_\br$ of unit length confined in a plane, whose orientation is 
specified by
the angle $\theta_\br$, with $\br=(i,j)$ being an integer vector  labelling 
the sites of a square lattice of size $N=N_x\times N_y$. It is known that this
model does not admit equations of motion and therefore its canonical dynamics
is usually  simulated either by Monte-Carlo methods~\cite{tobo,gupta1} or by
Langevin type equations~\cite{loft,jensen}. Microcanonical approaches
consist instead, either in considering a three component spin
model~\cite{evertz} or into adding a kinetic energy term~\cite{kogut}. The
latter method, which we follow in the  present work, can be also generalized to
other physical systems (see e.g. the application to lattice gauge
theories~\cite{calla}). All these different dynamics should display the same
static properties, as it has been  verified up to some
extent~\cite{kogut,leoncini}. We thus consider the Hamiltonian
\begin{equation}
{\cal H} \;= \;\sum_\br\frac{p_\br^2}{2}+ \sum_{\langle \br, \br' \rangle} 
[1-\cos(\theta_{\br'}-\theta_\br)]\;,
\label{hami}
\end{equation}
where $p_\br=\dot{\theta_\br}$ is the angular momentum of the rotator.  The sum
ranges over the four nearest neighbors of site $\br$, namely $\br'=\br\pm\ux$
and $\br'=\br\pm\uy$ where $\ux$ and $\uy$ are the  unit vectors parallel to
the lattice axis. It could be shown that (\ref{hami}) is obtained as the
classical limit of a quantum Heisenberg Hamiltonian with an anisotropy term
$\sum_{\br} (S_{\br}^z)^2$, using the representation introduced in
Ref.~\cite{V74}. We have set both the inertia of the rotators and the
ferromagnetic  coupling constant to unity so that the only physical control
parameter is  the energy per spin $e={\cal H}/N$. Actually, there is a second
constant  of the motion, the total angular momentum  $P = \sum_\br p_\br$,
whose choice affects the results in a trivial way. In the numerical simulations
we set $P=0$ to avoid global ballistic rotation. 

A previous study of the static properties of (\ref{hami})~\cite{leoncini} 
showed that the system undergoes a 
KTB transition~\cite{gulacsi} at $e=e_{KTB}\approx 1.0$, corresponding 
to a kinetic temperature $T_{KTB}\approx 0.89$, which is in agreement with
the value 0.894(5) obtained in the canonical ensemble~\cite{gupta1}.
One of the most striking features of the XY model is the presence of 
strong finite-size effects~\cite{bramwell}, e.g. the existence of a
sizable magnetization for large samples, despite the fact that
long-range order cannot occur for the infinite system. This has also 
some consequence on the dynamical correlation of the finite--size 
magnetization~\cite{LR01}. 

In the framework of linear-response theory, heat transport properties can be
analyzed by computing the autocorrelation  function (or, equivalently, the
power spectrum) of the total heat current vector ${\bf J}$ at equilibrium.  
We thus need a microscopic expression that can be worked out by the procedure 
followed for other similar models ~\cite{LLP03}. In brief, it amounts
to writing down a discretized continuity equation and, by means of the
equation of motion, identify the proper expression of the local flux in terms
of the canonical variables $(\theta_\br,p_\br)$. 
For model (\ref{hami}),  ${\bf J}=(J^{x},J^{y})$ can be written as a sum over
all lattice sites  
\begin{eqnarray}
J^{x}  \;=\; \frac12 \, \sum_{\br} \sin(\theta_{\br +\ux}-\theta_{\br})\,
\left[\dot\theta_{\br +\ux}+\dot\theta_{\br}\right] \\
J^{y}  \;=\; \frac12 \, \sum_{\br} \sin(\theta_{\br +\uy}-\theta_{\br})\,
\left[\dot\theta_{\br +\uy}+\dot\theta_{\br}\right] 
\label{flux}
\end{eqnarray}
This latter expression is the correct one in the microcanonical ensemble
with $P=0$. Incidentally, notice that a suitable counterterm should be 
subtracted out if one wishes to work in a different statistical ensemble 
\cite{G60}. 


The numerical integration of the equations of motion (with periodic boundary
conditions) is performed using the fourth-order McLahlan-Atela
algorithm~\cite{atela}, which is an explicit  scheme constructed from a suitable
truncation of the evolution operator that preserves the Hamiltonian structure.
One of the major merits of symplectic algorithms is that the error on the
energy does not increase with the length of the run. The chosen time step
(0.01-0.05 in our units) ensures that in every simulation energy fluctuates
around the prescribed value with a relative accuracy below $10^{-5}$.

As mentioned above, the main quantity of interest is the flux autocorrelation
function $\langle {\bf J}(t)\cdot {\bf J}(0)\rangle$. For numerical purposes,
we find more convenient to evaluate the power spectra $S(f)$, i.e. the squared
modulus of the Fourier transform of each component of ${\bf J}$, averaged over a
set of different initial conditions. These initial conditions were chosen 
by letting $\theta_\br=0$ and drawing the $\dot\theta_\br=0$ at random 
from a Gaussian distribution with zero average and unit variance. The 
momenta are then all rescaled by a suitable factor to yield the desired value
of the total energy. A transient is elapsed in order to start the averaging
from a more generic phase--space point. 

\section{Non--equilibrium simulations}

The non--equilibrium simulations have been performed by coupling all the rotators
on the left and right edges of the lattice with two thermal baths operating
at different temperatures $T_+$ and $T_-$. Periodic and fixed boundary
conditions have been adopted in the direction perpendicular ($y$) and parallel
($x$) to the thermal gradient, respectively.
Thermal baths have been simulated by applying the Nos\'e-Hoover method:  
\begin{eqnarray}
\ddot{\theta}_\br&=&-\frac{\partial V}{\partial \theta_\br}-
\dot{\theta_\br}\,\big[{\zeta_j}^{+} \delta_{i,1}+{\zeta_j}^{-} \delta_{i,N}\big]\\\nonumber
{\dot{\zeta}_{j}}^{+}&=&\frac{1}{\Theta_{+} ^2}\left( \frac{\dot{\theta}_{1,j}^2}{k_{B}T_{+}}-1\right)\\\nonumber
{\dot{\zeta}_{j}}^{-}&=&\frac{1}{\Theta_{-} ^2}\left( \frac{\dot{\theta}_{N,j}^2}{k_{B}T_{-}}-1\right)
\end{eqnarray}
Here, $V$ is the potential associated with (\ref{hami}), $\Theta_\pm$ are the
thermostats' response times, and $\delta$ is the usual Kronecker symbol. Notice
that each  rotator is thermostatted independently and, accordingly, the
Nos\'e-Hoover variables $\zeta^\pm$ are vectors of length $N_y$. For
computational purposes, simulations have been performed  by fixing the aspect
ratio  $R = N_y/N_x < 1$. The choice of $R$ results from a trade--off between
minimizing the number of rotators ($R$ small) and dealing  with a genuinely
$2d$ lattice ($R\sim 1$). Indeed, too small values of $R$ would require
considering larger system sizes to clearly observe $2d$ features. For small
lattices, we checked that the results are almost independent of the ratios
employed hereby. 

Since we are interested in the average values of the flux we have 
to check that the non--equilibrium stationary state is indeed attained. To this
aim, we monitored that the average fluxes towards the boundaries
\begin{eqnarray}
J^{+}&=& -\sum_j \zeta_j^{+} \dot{\theta}_{1,j}^2\\\nonumber
J^{-}&=& -\sum_j \zeta_j^{-} \dot{\theta}_{N,j}^2
\end{eqnarray}
were equal to the flux in the bulk, namely $\overline{J^\pm}=\overline{J^x}$
(the overline denotes a time average henceforth).
 
As observed before~\cite{LLP03}, the choice of the thermostat response times is
crucial to the time needed to reach the stationary state, and to control the
values of thermal resistance at the boundaries. In order to fasten the
convergence, the initial conditions have been generated by thermostatting each
particle to yield a linear temperature profile along the $x$ direction. This
method is very efficient, especially for large lattices, when 
thermalization within the bulk may be significantly slow.

Once the steady non--equilibrium state is attained, the relevant observables are
accumulated and averaged in time. In particular we evaluated the local kinetic
temperature $k_B T_\br \; =\; \overline{\dot\theta_\br^2}$ and  the average
energy fluxes $\overline{J^x}$ and $\overline{J^y}$, as defined  by formulae
(\ref{flux}), as well as the average fluxes towards the reservoirs
$\overline{J^\pm}$. For obvious simmetry reasons, we expect (and indeed found)
the $\overline{J^y}$ vanishes up to the statistical accuracy. Moreover, 
$T_\br$ depends only on the location along the $x$ axis, and a
further average is performed along $y$. Once the the average flux is 
computed, we evaluated the thermal conductivity coefficient from 
Fourier's law as 
\begin{equation}
\kappa (N_{x})\;=\;\frac{\overline{J^x}}{\vert \nabla
T\vert}\;\simeq\;\frac{ \overline{J^x} N_{x}}{T_+-T_-}
\label{kappa}
\end{equation}
The last equality is only approximate, since the actual thermal gradient within
the lattice is usually smaller than $(T_+-T_-)/N_x$, due to boundary resistance
effects~\cite{LLP03}. In other words, the coefficient evaluated in this way 
should be regarded as an effective conductivity including both boundary and
bulk scattering. As we are going to show in the next section the rescaled
stationary temperature profiles, obtained for different values of $N_x$
are such that the temperature gradient actually scales like $N_x^{-1}$.
Therefore, definition~(\ref{kappa}) yields the same scaling behavior of 
the bulk conductivity.

\section{The disordered phase}

\begin{figure}[!!here]
\centering\includegraphics[clip,height=7cm]{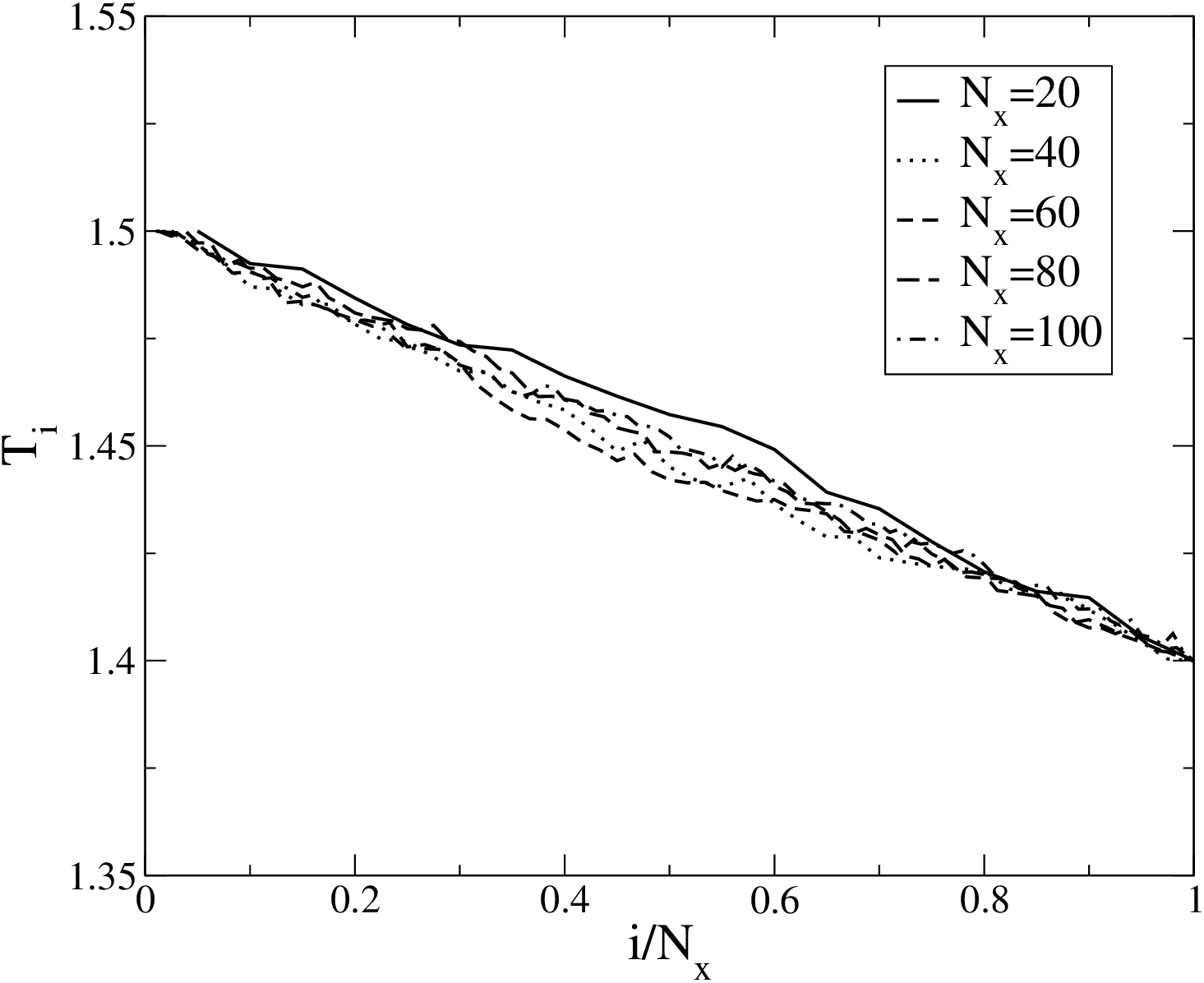}
\caption{The temperature profile in the high--temperature phase
for different values of $N_x$, which have been rescaled to
the unit length. The response times of the Nos\`e--Hoover thermostats are 
$\Theta_\pm = 1.0$, which guarantee negligible boundary impedance effects.
}
\label{proft1}
\end{figure}

\begin{figure}[!!here]
\centering\includegraphics[clip,height=7cm]{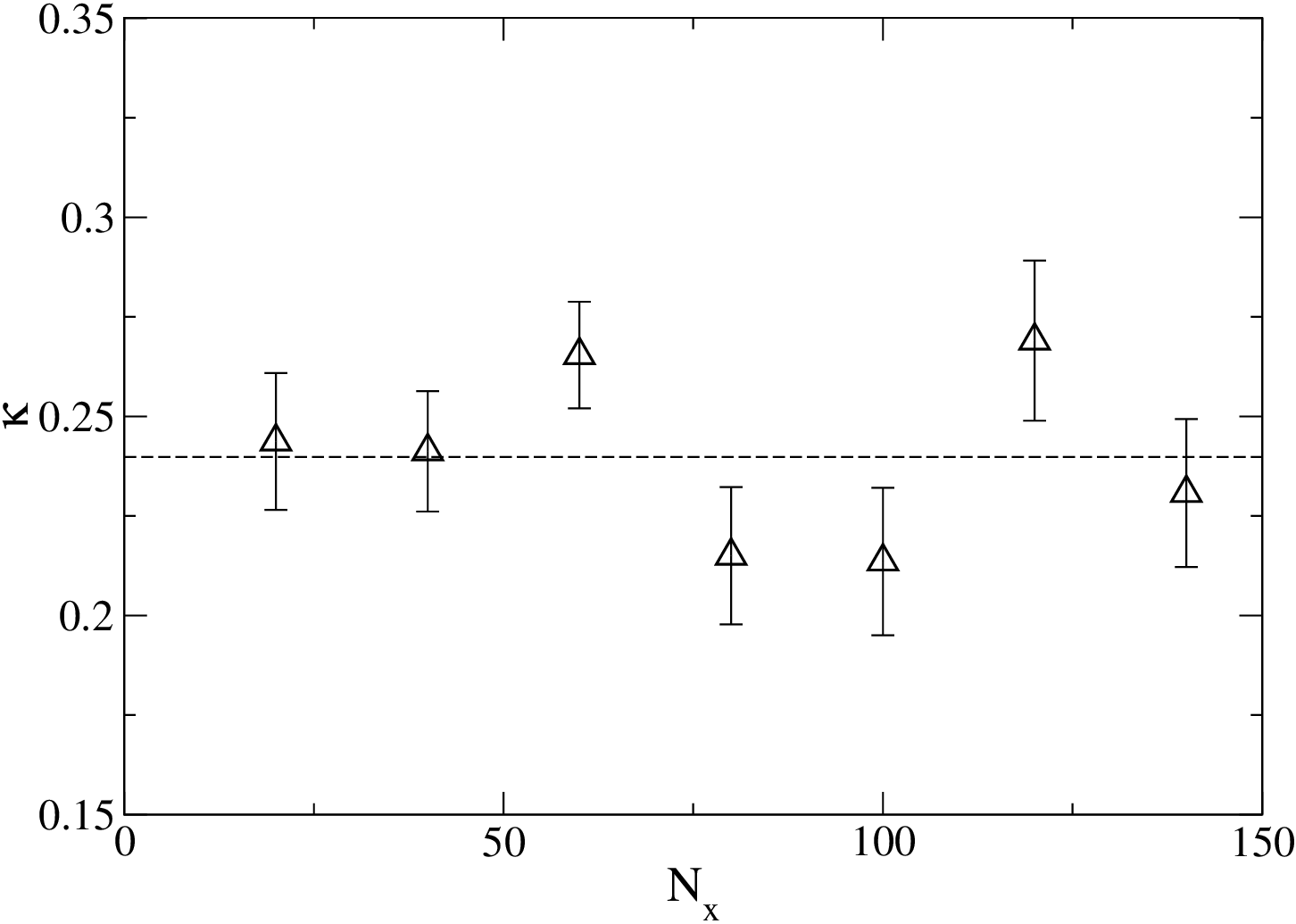}
\caption{Finite-size conductivity in the disordered phase.
Nos\`e--Hoover thermostat with response times fixed to $1.0$;
Each point results from an average of 16 independent 
runs of about $10^{6}$ time units each. The
error bars are the error on the mean and the horizontal line 
is the average of all the measured values, $\kappa=0.24$.}
\label{kappa1}
\end{figure}

Let us start discussing the case of high energies or temperatures where the
system is away from criticality. In the non--equilibrium simulations we fixed
$T_+=1.5$ and $T_-=1.4$ which are both well above the KTB transtion
temperature.  Both response times of the Nos\'e--Hoover thermostats have been
set to the same value, $\Theta_\pm = 1.0$. In fact, we have found
empirically that such a choice minimizes boundary impedance  effects,thus
allowing for larger values of the heat flux.  In order to improve the
statistics, the temperature profiles and the measures of the heat flux have
been averaged over 16  independent initial conditions. Numerical simulations
have been performed by fixing the value of the apsect--ratio to $R=1/2$ and by
increasing $N_x$ up to 140. Some of the temperature profiles are reported in
Fig.~\ref{proft1}. They all exhibit a linear shape, which testifies to the
expected temperature profile when Fourier's law holds. In Fig.~\ref{kappa1} we
show that the thermal  conductivity, as defined by (\ref{kappa}), is independent
of $N_x$, with fluctuations around the average value, extending up to
some 10 \%. 

\begin{figure}[here]
\vskip 1.5cm
\centering
\includegraphics[clip,height=8cm]{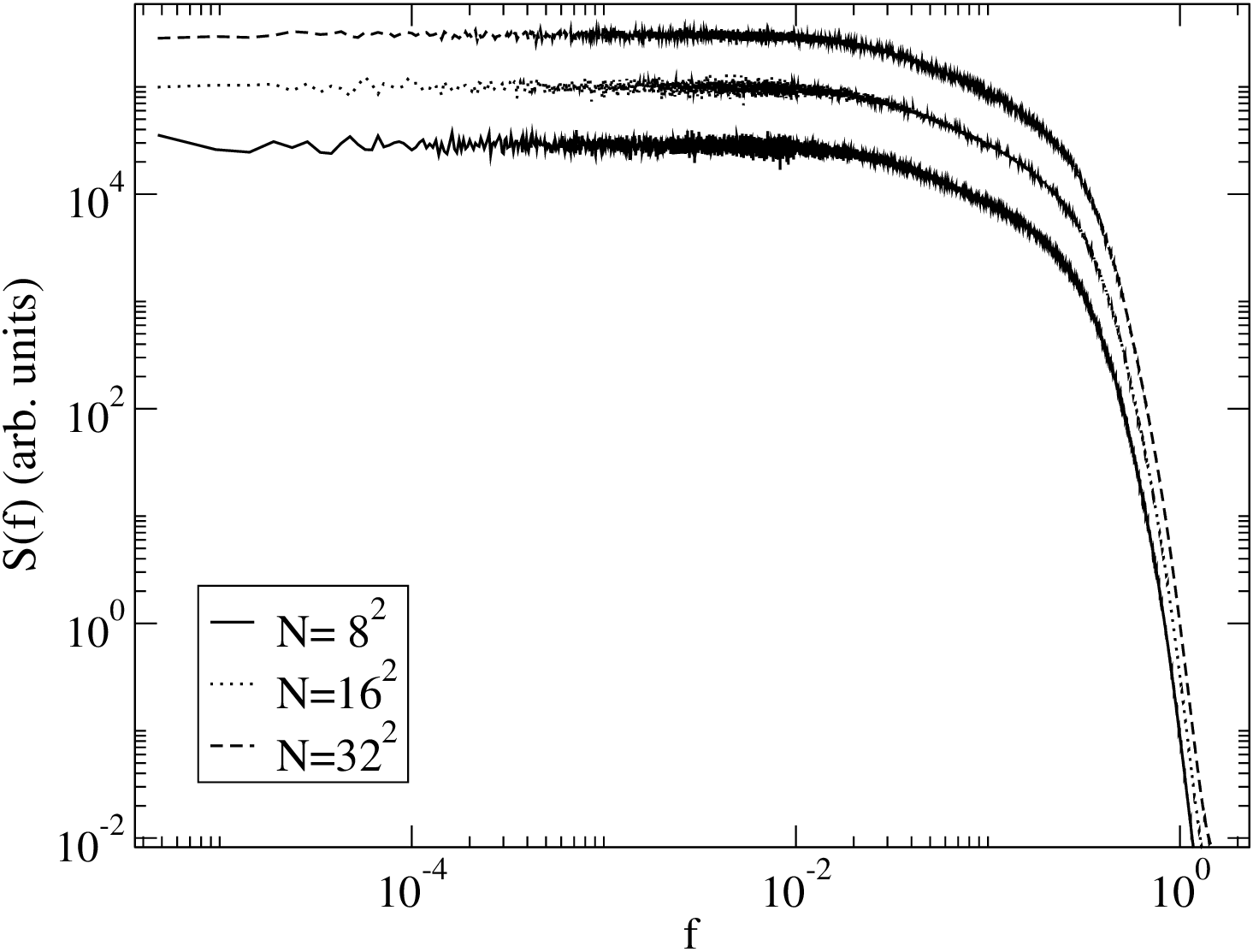}
\caption{
Power spectra of the heat current in the disordered phase
for three different lattice sizes, $N = 8^2, 16^2, 32^2$. 
Data are averaged over 200 random initial conditions. In order to
minimize statistical fluctuations, a further averaging of the data over
contiguous frequency intervals has been performed. The three
spectra actually almost overlap: this is why we have presented 
them after a vertical arbitary shift in order to
better distinguish one from each other.}
\label{disspe}
\end{figure}

These results have been compared with those obtained from equilibrium
simulations. The value of the energy density has been chosen $e=2$. Actually,
this value corresponds approximately  to the average temperature, $(T_+ +
T_-)/2$, of the above mentioned non--equilibrium simulations. We want to point
out that, for very large values of $e$, the  kinetic energy dominates over the
potential one and the system approaches  the integrable limit of independent
free rotators. Accordingly, in this limit the lattice is expected to behave as
a perfect insulator, since  the time scale for transmitting any energy
fluctuation diverges. In this respect, the choice $e=2$ is appropriate, also
because one can  observe convergence of the quantities of interest over
reasonable  simulation times (typically, $10^6$ time units). In
Fig.~\ref{disspe} we show the heat flux spectra: they are independent of the
lattice sizes and  tend to a constant for small frequencies. This in a clear
confirmation that no long--time tail is detectable and the thermal conductivity
is a  well--defined quantit in the thermodynamic limit.  It should be remarked
that the lineshape of these spectra cannot be fitted  by a simple Lorenzian.
This implies that in the high--temperture phase the decay of time--correlations
cannot be reduced to a simple exponential law. 

\section{The critical phase}

A more interesting situation appears in the low--temperature phase. For what
concerns the non--equilibrium calculations, we have fixed the thermostats'
temperatures $T_\pm$ to be well below the KTB transition value. At variance
with the high-temperature phase, here the values of the response time of the
Nos\'e--Hoover  thermostats have to be properly tuned in order to minimize
boundary  impedence effects. In particular, we have determined empirically the
values $\Theta_+=2.0$ and $\Theta_-=6.0$. Such different values, between
themselves and also with respect to the high--temperature phase, indicate that
fluctuations have to be slowed-down significantly in order to cope with the
typical time scales of the dynamics. Moreover, we have performed the same
statistical averaging as in the high--temperature case. 

\begin{figure}[!!here]
\centering\includegraphics[clip,height=7cm]{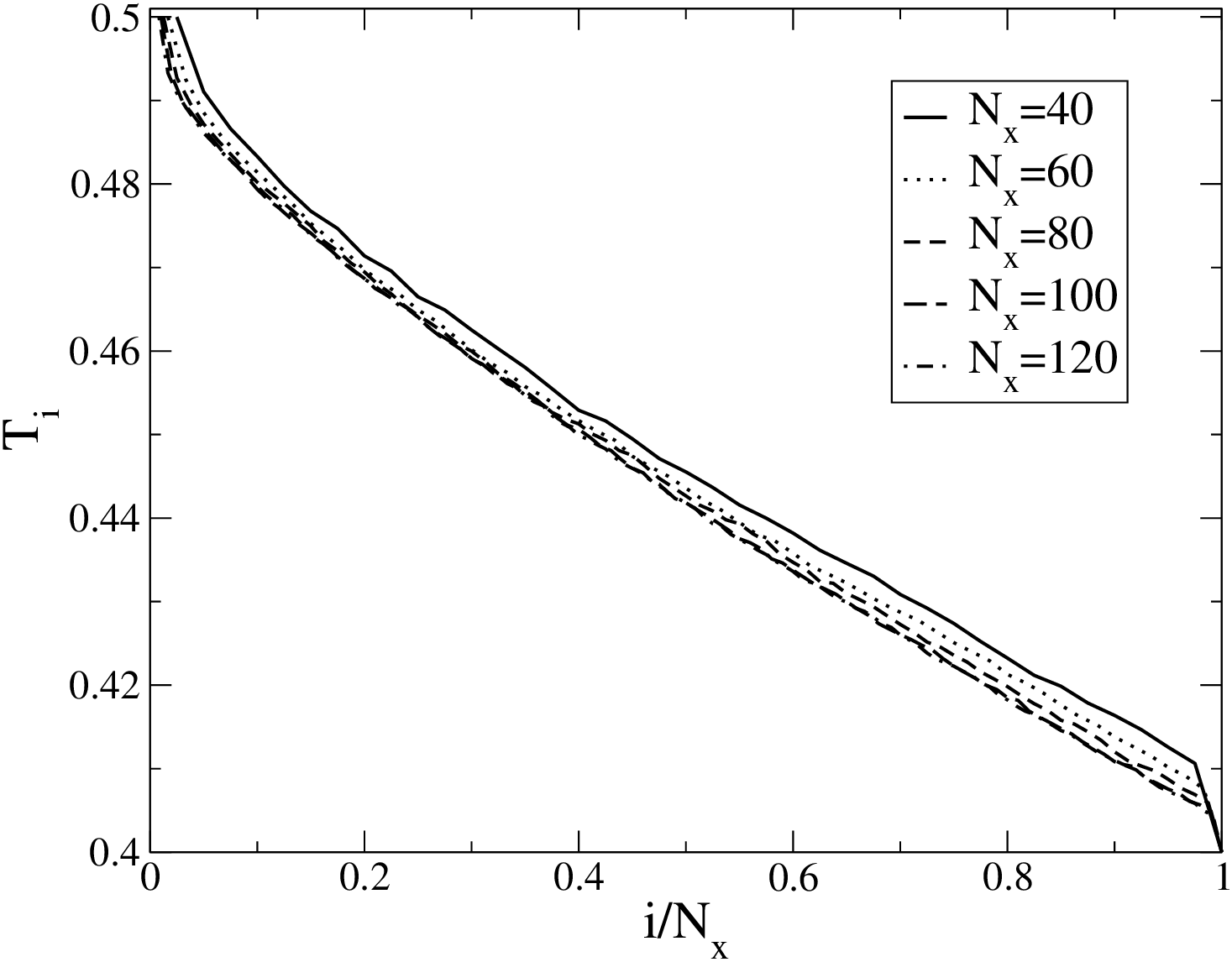}
\caption{The temperature profile in the low--temperature phase
for different values of $N_x$, which have been rescaled to
the unit length.
The response times of the Nos\`e--Hoover thermostats are 
$\Theta_+ = 2.0$ and $\Theta_- = 6.0$, which minimize boundary
impedence effects.}
\label{proft2}
\end{figure}

In Fig.~\ref{proft2} we show the temperature profiles obtained for
$T_+=0.5, T_-=0.4$ and  different values of $N_x$. They exhibit a good data
collapse when $N_x$ is rescaled to the unit length. This confirms that also in
the low--temperature phase the thermal gradient scales  like the inverse of the
system size, $\nabla T \sim (T_+ - T_-)/N_{x} $. On the other hand, the
temperature profile has assumed the typical non--linear shape, which testifies
to anomalous thermal conductivity. Moreover, this shape is similar to the
temperature profiles of the one and two--dimensional Fermi--Pasta--Ulam model
\cite{LLP03}.

\begin{figure}[!!here]
\vskip 1.5cm
\centering
\includegraphics[clip,height=7cm]{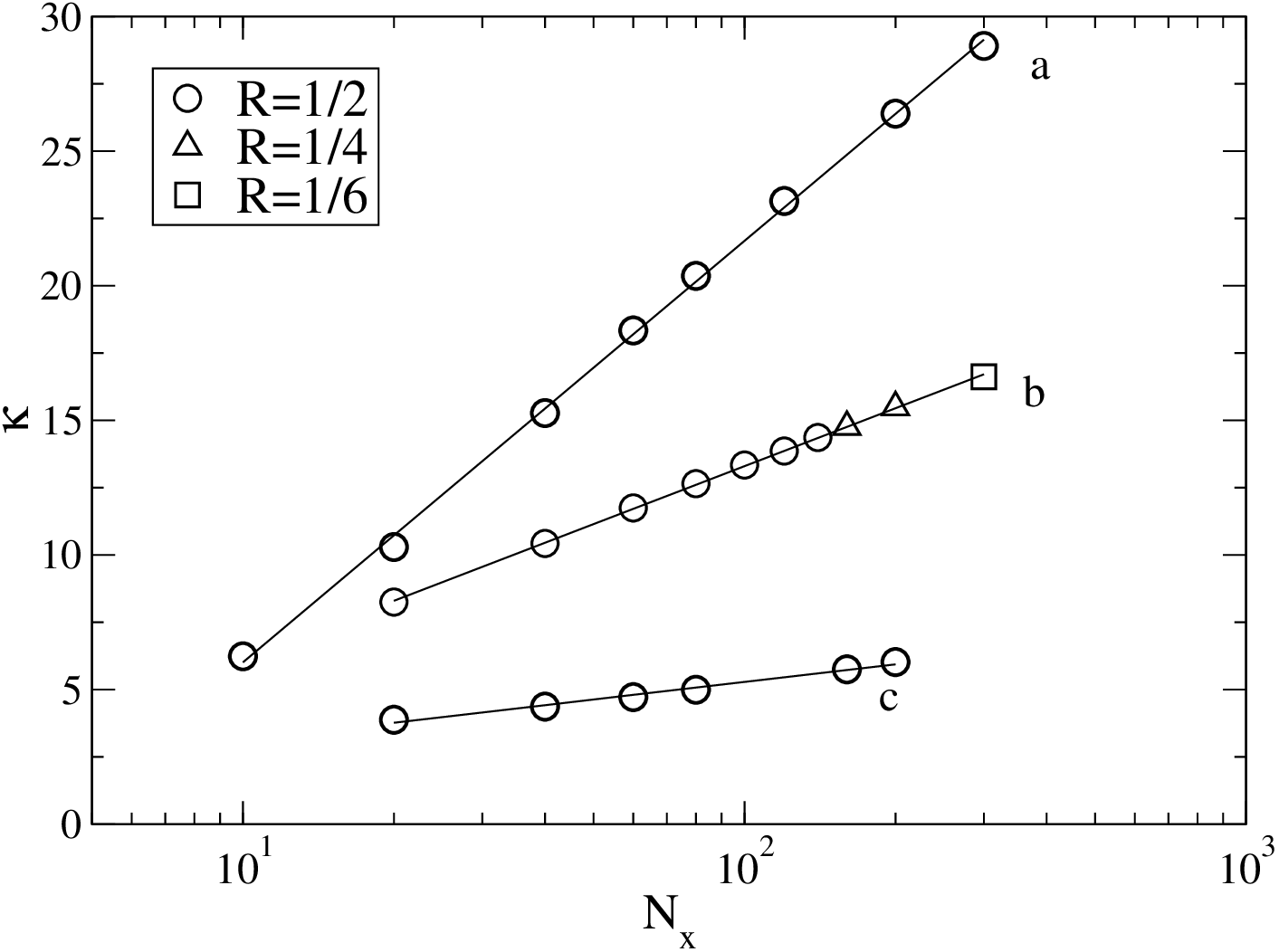}
\label{kbelow}
\caption{
The thermal conductivity $\kappa$ versus $N_x$ in the low--temperature phase
for $T_-=0.3, T_+=0.4$ (a), $T_-=0.4, T_+=0.5$ (b) and  $T_-=0.6, T_+=0.7$ (c).
Each point is the result of an average over 16 independent  initial conditions,
each one lasting for $10^{6}$ time units. The error bars are of the order of
the symbol's size. The solid lines are a best--fit with a logarithmic law, 
Eq.~(\ref{loga}).}
\end{figure}

The finite-size thermal conductivity  as a function of the longitudinal size
$N_x$  is reported in Fig.~\ref{kbelow} for three different values of  the
boundary temperatures (with $T_+-T_-$ kept fixed to 0.1).  As expected,
increasing $T_\pm$ the conductivity decrease.  More importantly, for fixed
temperatures, the data all exhibit a systematic increase with $N_x$.  In
analogy with what found in the $2d$ Fermi-Pasta-Ulam model~\cite{LL00},  the
data for lower temperatures (curves (a) and (b)) can be well fitted  by a
logarithmic law
\begin{equation}
\kappa(N_x) \;=\; C_1 \,+\, C_2 \ln N_x \,\, .
\label{loga}
\end{equation}
Actually, a closer inspection of data set (c) reveals that the logarithmic
fit is rather poor. In view also of the limited size range 
we have been able to explore, a convincing 
estimate of the growth law is unfeasible with the data at hand. 
This is presumably due to the fact that approaching $T_{KTB}$ 
requires longer times and sizes. In fact, we have observed that
convergence of the averages considerably slows down in this 
region (about a factor 5 in passing from simulation (b) to (c)).
The effect may be caused by slow thermalization of vortex pairs.  
Indeed, in this temperature
region the vorticity starts to become sizeable \cite{leoncini} 
and a slower kinetics as well as relevant correction to scaling 
may thus be expected.

Following the analysis performed for the high--temperature phase, we have
performed also microcanonical simulations, with periodic boundary conditions
imposed in both lattice directions. We only considered the energy density
$e=0.5$ which roughly corresponds to $T=0.45$. 

\begin{figure}[!!here]
\vskip 1cm
\centering
\includegraphics[clip,height=7cm]{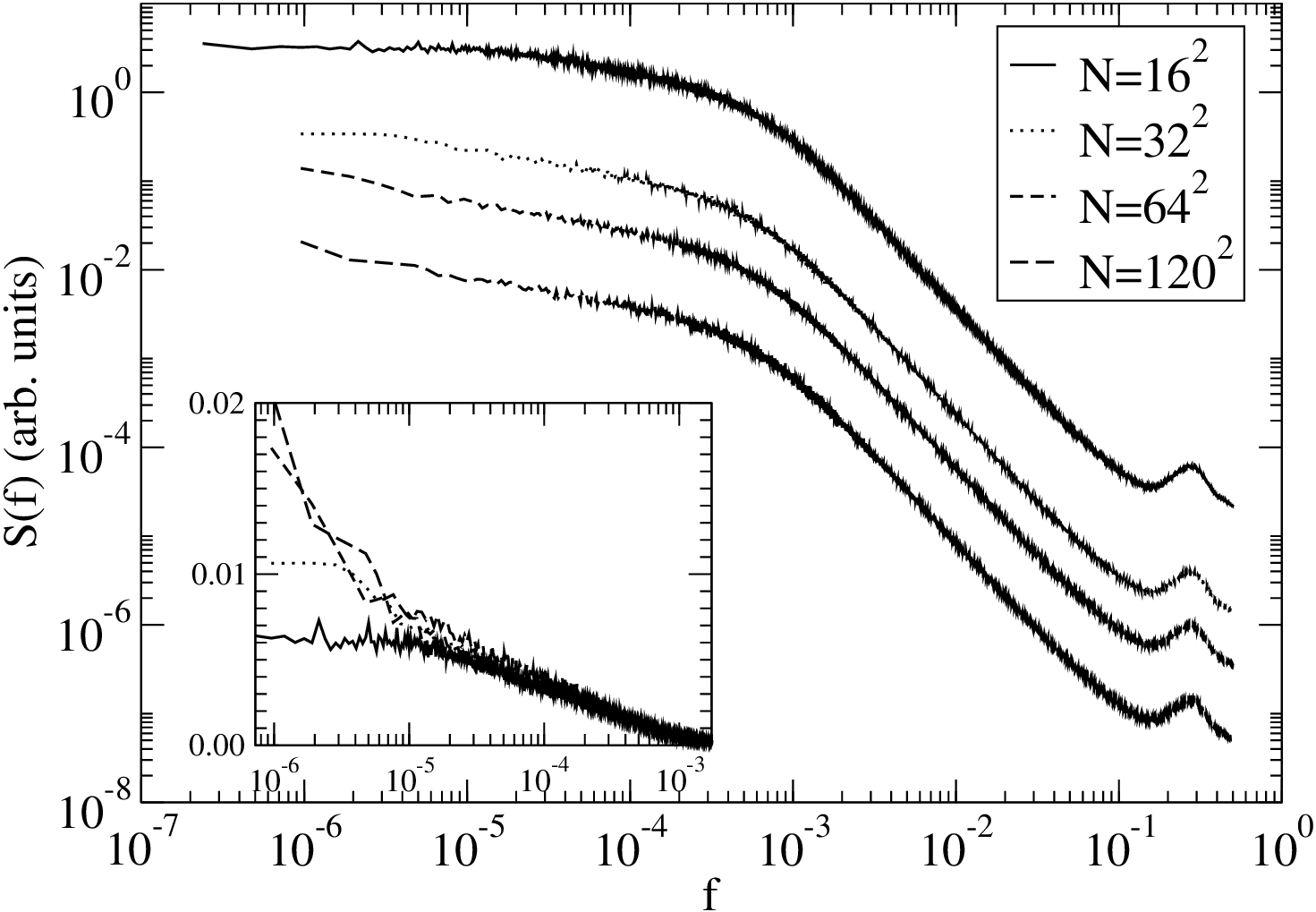}
\caption{
Power spectra of the heat current in the low--temperature phase ($e=0.5$).
for four different lattice sizes $N=16^2, 32^2, 64^2, 120^2$. 
Data are averaged over 400 random initial conditions. In order
to minimize statistical fluctuations, a further averaging of the data over
contiguous frequency intervals has been performed. 
The spectra are presented after a vertical arbitrary shift 
to better distinguish one from each other. The inset is an enlargement
of the low--frequency region in log--lin scale.}
\label{speclow}
\end{figure}

In Fig.~\ref{speclow}, we report the power spectra of the heat flux for
different lattice sizes. For large frequencies ($f > f_c \approx 10^{-3}$) we
have a $f^{-2}$ behaviour which suggests a fast (exponential) decay of the
correlation at  short times. The crossover frequency $f_c$ should be related to
some typical time--scale for the hydrodynamic effects to set in. In the
low--frequency limit, $f<f_c$, the data are consistent with a logarithmic 
singularity of the type (see the inset of Fig.~\ref{speclow}) 
\begin{equation} 
S(f) \;=\; A \,+\,B \ln f 
\label{specfr} 
\end{equation} 
which, in turn, corresponds a $t^{-1}$ tail of the autocorrelation function.
According to the argument exposed below Eq.~(\ref{anomal}), 
this would yield the logarithmic divergence (see Eq.(\ref{loga})). 
Altogether, we conclude that the equilibrium and nonequilibrium aproaches
yield the same divergent behaviour.

It must be admitted that the fitting of the low--frequency part with formula
(\ref{specfr}) is convincing only for the larger sizes.  Actually, a power-law
fit $S(f) \propto f^{-0.4}$ is also compatible with the data obtained for
smaller $N$ values. On the other hand, this estimate cannot be taken seriously
for a twofold reason. First of all, it may be easily attributed to finite-size
effects. Moreover, it would imply a power--law divergent conductivity which, in
turn, would be inconsistent with the non--equilibrium data (see again curve (b)
in  Fig.~\ref{kbelow}). 

In view of the above fact, one may wonder why equilibrium simulations should be
much more sensitive to finite-size effects than nonequilibrium ones. A
reasonable qualitative explanation goes as follows. At low energies, where
the isolated system is almost harmonic, it is very unlikely that a rotator
turns from small oscillations to fast rotations. Moreover, for a small
size system this process is even more unlikely as it demands a sufficiently
large local energy fluctuation. Since this is the main
scattering mechanism, one should wait for very long simulations before 
recovering the true asymptotic regime of energy transport. 
Conversely, in nonequilibrium simulations thermal baths act as 
external sources of fluctuations. These may favour the creation of the 
nonlinear excitations, thus shortening the time scale needed for the effects 
of the scattering process to be appreciated.



\section{Concluding remarks}

We have found numerical evidence that transport properties of the XY model  on
a finite lattice are drastically different in the high--temperature and in the
low--temperature phases. In particular, thermal conductivity is finite in the
former case, while in the latter it does not converge up to lattice sizes of
order 10$^4$. In the region where vorticity is negligible ($T < 0.5$) the
available data suggest a logarithmic divergence with the system size analogous
to the one observed for coupled oscillators \cite{LL00}. Close to $T_{KBT}$,
where a sizeable density of bounded vortex pairs are thermally excited, our 
data still suggest a divergence, whose law we cannot reliaby estimate.

We want to point out that these results have been obtained consistently for 
equilibrium and nonequilibrium simulations. The equivalence between these
approaches is not granted a priori. On the other hand, this property has been
verified for many other similar models \cite{LLP03}, whose dynamics is of 
Hamiltonian type. In this respect, different choices of the dynamics (e.g. 
Monte--Carlo) may not necessarily lead to the same conclusion.

This is a very important and interesting physical result: it indicates that in
the low--temperature phase some materials or states of matter, may behave as
anomalously efficient heat conductors. For instance, as a direct consequence of
the studies performed in this paper on the $2d$ XY model, liquid Helium films
should be included in this class of materials. An experimental test confirming
the prediction of the logarithmic divergence of the heat conductivity with the
system size would be highly appropriate and welcome.

A complete hydrodynamic theory of  the $2d$ XY model could certainly help in
clarifying many of the aspects that our numerical approach cannot fully
assess.  For instance, an approach based on spin--waves proved out
to be be effective in evaluating the dynamical correlations of the
low-temperature phase \cite{nelson}. On the other hand, estimates of the 
energy--current correlators may be technically more difficult. Indeed, the heat
flux is a constant to leading order and, accordingly, one has to account for
higher--order terms for the theory to make any sense. In this
respect the calculations should be conceptually equivalent to estimating
spin--waves lifetimes \cite{wysin}. It is not however clear how to include the
effects of vortices and finite--size magnetization in this framework. The
hydrodynamics in the high--temperature phase presumably should go through less
technical troubles, although the construction of, say, a large--deviation
functional is far from trivial. Anyway, an effort in this direction is in our
future agenda.

\ack

This work is supported by the INFM-PAIS project {\it Transport
phenomena in low-dimensional structures} and  is part of the PRIN2003 project
{\it Order and chaos in nonlinear extended systems} funded by MIUR-Italy.
Part of the numerical simulation were performed at CINECA supercomputing
facility through the INFM {\it Iniziativa trasversale ``Calcolo Parallelo"}
entitled  {\it Simulating energy transport in low-dimensional systems}.


\section*{References}

\begin{thebibliography}{99} 


\bibitem{PR75} Pomeau Y and R\'esibois R, 1975 
{\it Phys. Rep.} {\bf 63} 19

\bibitem{kirk} Kirkpatrick TR, Belitz D and Sengers JV, 2002
{\it J. Stat. Phys.} {\bf 109}, 373 

\bibitem{LLP97} Lepri S, Livi R and Politi A, 1997 
{\it Phys. Rev. Lett.} {\bf 78} 1896

\bibitem{LLP03} Lepri S, Livi R and Politi A, 2003
{\it Phys. Rep.} {\bf 377} 1 

\bibitem{NR02}  Narayan O and Ramaswamy S, 2002
{\it Phys. Rev. Lett.} {\bf 89}  200601 

\bibitem{LLP03b} Lepri S, Livi R and Politi A, 2003
{\it Phys. Rev. E} {\bf 68} 067102 

\bibitem{D01} Dhar A, 2001 
{\it Phys. Rev. Lett.} {\bf 86} 3554; 

\bibitem{CP03} Casati G and Prosen T, 2003
{\it Phys. Rev. E} {\bf 67},  015203

\bibitem{LL00} Lippi A and Livi R, 2000
{\it J. Stat. Phys.} {\bf 100}, 1147 

\bibitem{GY02} Grassberger P and Yang L, 
unpublished [cond-mat/0204247]

\bibitem{saito} Saito K, Takesue S and Miyashita S, 1999
{\it Phys. Rev. E} {\bf 59} 2783

\bibitem{Hu} Hu B, Li B and Zhao H, 2000 
{\it Phys. Rev. E} {\bf 61} 3828 ; 
Aoki K and Kusnezov D, 2000
{\it  Phys. Lett. A} {\bf 265} 250 

\bibitem{gulacsi} Gul\'acsi Z and Gul\'acsi M, 1998
{\it Adv. Phys.} {\bf 47} 1 

\bibitem{GLPV00} Giardin\`a C {\it et al.}, 2000
{\it Phys. Rev. Lett.} {\bf 84 } 2144;
Gendelmann O V, and Savin A V, 2000 
{\it ibid.} {\bf 84 } 2381

\bibitem{DP97} Dellago C and Posch H A, 1997
{\it Physica A} {\bf 237} 95.



\bibitem{tobo} Tobochnik J and Chester G V, 1979
{\it Phys. Rev. B} {\bf 20} 3761

\bibitem{gupta1} Gupta R and Baille C F, 1992
{\it Phys. Rev. B} {\bf 45} 2883

\bibitem{loft} Loft R and DeGrand T A, 1987
{\it Phys. Rev. B} {\bf 35} 1987

\bibitem{jensen} Jensen L M, Kim B J and Minnhagen P, 2000
{\it Phys. Rev. B} {\bf 61} 15412

\bibitem{evertz} Evertz H G and Landau D P, 1996
{\it Phys. Rev. B} {\bf 54} 12302
 
\bibitem{kogut} Kogut J and Polonyi J, 1986
{\it Nucl. Phys. B} [FS15] {\bf 265} 313

\bibitem{calla} Callaway D J E  and Rahman A, 1982
{\it Phys. Rev. Lett.} {\bf 49} 613

\bibitem{leoncini} Leoncini X, Verga A D and Ruffo S, 1998
{\it Phys. Rev. E} {\bf 57} 6377

\bibitem{V74} J. Villain, 1974 
{\it J. Phys. (Paris)} {\bf 35} 27

\bibitem{bramwell} Bramwell S T and Holdsworth P C W, 1993
{\it J. Phys. Condens. Matter} {\bf 5} L53  

\bibitem{LR01} Lepri S and Ruffo S, 2001
{\it Europhys. Lett.} {\bf 55 } 512  

\bibitem{G60} Green M S, 1960
{\it Phys. Rev.} {\bf 119},  829 

\bibitem{atela} McLachlan P I and Atela P, 1992
{\it Nonlinearity} {\bf  5} 541

 \bibitem{nelson} Nelson D R and Fisher D S, 1977
{\it Phys. Rev. B} {\bf 16} 4945

\bibitem{wysin} Wysin G M, Gouvea M E, and Pires A S T, 2000
{\it Phys. Rev. B} {\bf 62} 11585 


\end{thebibliography}
\end{document}